\documentclass{article}
\usepackage{spconf,amsmath,graphicx}
\usepackage{booktabs}

\usepackage{url, times, booktabs, tabularx, hyperref}
\usepackage{multirow}

\usepackage{tikz}

\title{A BENCHMARK OF STATE-OF-THE-ART SOUND EVENT DETECTION SYSTEMS EVALUATED ON SYNTHETIC SOUNDSCAPES}
\name{Francesca Ronchini$^{1}$,
       Romain Serizel$^{1}$
       \thanks{This work was made with the support of the French National Research Agency, in the framework of the project LEAUDS Learning to understand audio scenes (ANR-18-CE23-0020), the project CPS4EU Cyber Physical Systems for Europe (Grant Agreement number: 826276) and the French region Grand-Est. Experiments presented in this paper were carried out using the Grid5000 testbed, supported by a scientific interest group hosted by Inria and including CNRS, RENATER and several Universities as well as other organizations (see \href{https://www.grid5000}{https://www.grid5000}).}
       \thanks{The authors would like to thank all the other organizers of DCASE 2021 Challenge Task 4.}
}
\address{$^1$Université de Lorraine, CNRS, Inria, Loria, Nancy, France \\
  }
  
\begin{document}
\ninept

\maketitle
\begin{abstract}
This paper proposes a benchmark of submissions to Detection and Classification Acoustic Scene and Events 2021 Challenge (DCASE) Task 4 representing a sampling of the state-of-the-art in Sound Event Detection task. The submissions are evaluated according to the two polyphonic sound detection score scenarios proposed for the DCASE 2021 Challenge Task 4, which allow to make an analysis on whether submissions are designed to perform fine-grained temporal segmentation, coarse-grained temporal segmentation, or have been designed to be polyvalent on the scenarios proposed.

We study the solutions proposed by participants to analyze their robustness to varying level target to non-target signal-to-noise ratio and to temporal localization of target sound events. A last experiment is proposed in order to study the impact of non-target events on systems outputs. Results show that systems adapted to provide coarse segmentation outputs are more robust to different target to non-target signal-to-noise ratio and, with the help of specific data augmentation methods, they are more robust to time localization of the original event. Results of the last experiment display that systems tend to spuriously predict short events when non-target events are present. This is particularly true for systems that are tailored to have a fine segmentation. 

\end{abstract}

\section{Introduction}
\label{sec:intro}
The task of Sound Event Detection (SED) consists in correctly detecting target sound events present in an audio clip.  SED systems are expected to produce strongly-labeled outputs (i.e.~detect sound events with a start time, end time, and sound class label)~\cite{virtanen2018computational}. Multiple events can be present in each audio recording, including overlapping target sound events and potentially non-target sound events.

Since 2018, DCASE Challenge Task 4 proposes to address the SED problem in a context where systems are provided with unlabeled and weakly labeled recorded clips (without any timing information) for training~\cite{serizel2018_DCASE}. In 2019, we proposed to use an additional training set composed of strongly labeled synthetic soundscapes~\cite{Turpault2019_DCASE}. These are cheap to obtain but they can introduce a domain mismatch when SED systems have to operate on recorded clips. 

Task 4 of the DCASE Challenge has been motivated by the fact that manually obtaining strong annotations is time consuming and therefore expensive. Additionally, strong annotations are known to be error-prone, mainly because of the subjective judgement of each annotator regarding onsets and offsets localization. A large set of annotations have been recently released on a portion of Audioset~\cite{hershey2021benefit}. Doing this for each specific application is not feasible and strong labels data remain subject to human annotators interpretations. Therefore, it is still relevant to explore to what extent it is possible to cheaply train a SED system from an heterogeneous dataset~\cite{turpault:hal-02891665}.

An additional advantage of using synthetically generated soundscapes is the possibility to have full control over the properties of the soundscapes~\cite{salamon2017scaper}. This allows to generate custom synthetic clips to untangle some of the many problems faced by a SED system operating under real conditions. Since 2019, within DCASE Task 4, 
synthetic soundscapes have been designed to target some specific SED open problems and they have been proposed in the evaluation set to the challenge participants in order to obtain a benchmark of state-of-the-art SED systems on these specific problems~\cite{Serizel2020_ICASSP,turpault2021sound}.

During previous iterations of the challenge, we have observed that obtaining 
an accurate time segmentation was one of the most prominent challenges of the SED systems~\cite{ turpault2021sound,serizel:hal-02114652,serizel:hal-02355573}. In 2021, we proposed to change the metric from an event-based F-score~\cite{mesaros_metrics_2016} to a polyphonic sound detection score (PSDS)~\cite{bilen2020framework}. 
This metric has been proven to be more robust in labeling subjectivity than the collar-based match metric (such as event-based F-Score) and it allows focusing on specific application scenarios~\cite{ferroni:hal-02978422}. In particular depending on how finely sound events need to be localized in time, a specific set of parameters can be used.

\begin{table*}[t!]
\centering
 \begin{tabular}{l|l|c|c|l|c|c}
  \toprule
  Ref&Submission code system 1 & PSDS\_1 & PSDS\_2 & Submission code system 2 & PSDS1 & PSDS2 \\
  \midrule 
  \cite{Zheng2021}&Zheng\_USTC\_task4\_SED\_1 & 0.45 & 0.67 
  & Zheng\_USTC\_task4\_SED\_3 & 0.39 & 0.75 \\
  
  \cite{Kim2021}&Kim\_AiTeR\_GIST\_SED\_4 & 0.44 & 0.67 & 
  Kim\_AiTeR\_GIST\_SED\_4 & 0.44 & 0.67 \\
  
  \cite{Lu2021}&lu\_kwai\_task4\_SED\_1 & 0.42 & 0.66 & 
  lu\_kwai\_task4\_SED\_3 & 0.15 & 0.69\\
  
  \cite{Nam2021}&Nam\_KAIST\_task4\_SED\_2 & 0.40 & 0.61 & 
  Nam\_KAIST\_task4\_SED\_4 & 0.06 & 0.72\\
  
  \cite{Ebbers2021}&Ebbers\_UPB\_task4\_SED\_3 & 0.42 & 0.64 & 
  Ebbers\_UPB\_task4\_SED\_4 & 0.36 & 0.64 \\
  
  \cite{Tian2021}&Tian\_ICT\_TOSHIBA\_task4\_SED\_1 & 0.41 & 0.59 & Tian\_ICT\_TOSHIBA\_task4\_SED\_1 & 0.41 & 0.59\\
  
  \cite{Gong2021}&Gong\_TAL\_task4\_SED\_3 & 0.37 & 0.63 & 
  Gong\_TAL\_task4\_SED\_3 & 0.37 & 0.63 \\
  
  \cite{Wang2021}&Wang\_NSYSU\_task4\_SED\_3 & 0.34 & 0.65 & 
  Wang\_NSYSU\_task4\_SED\_4 & 0.30 & 0.66 \\
  
  \cite{Yu2021}&Cai\_SMALLRICE\_task4\_SED\_2 & 0.37 & 0.58 & 
  Cai\_SMALLRICE\_task4\_SED\_3 & 0.37 & 0.60 \\
  
  \midrule
  
  \cite{ronchini2021impact}& Baseline & 0.31 & 0.55 & 
  Baseline & 0.31 & 0.55 \\ 
 \bottomrule
 \end{tabular}
 \caption{PSDS\_1 and PSDS\_2 of the nine highest-ranked teams based on rankings score plus the baseline.}
 \label{tab:rank}
\end{table*} 

This paper benchmarks DCASE 2021 Challenge Task 4 submissions which represent a sample of the state-of-the-art in SED. We investigate the solutions proposed by participants depending on time segmentation constraints, analyse the robustness of the different submitted systems to varying levels of target to non-target signal-to-noise ratio (TNTSNR)
and a varying time localization of target sound events. The SED submissions are evaluated according to the two PSDS~\cite{PSDS:2020} scenarios defined for the DCASE 2021 Challenge Task 4. \footnote{To promote reproducibility, the mapping files, the ground-truth of the proposed subsets, and the submissions of the teams are made available under an open-source license \url{https://zenodo.org/record/5949149}}.

\section{Datasets and task setup}
\label{sec:dataset}

\subsection{Task setup and evaluation metrics}
\label{subsec:setup}


In order to better understand the expected behaviour of each submission and with the aim to emphasize different systems properties for the two scenarios considered in the DCASE 2021 Challenge Task 4, the following definitions are given.

\textbf{Scenario 1:}
The system needs to react fast upon an event detection (e.g. to trigger an alarm, adapt home automation system \dots). The localization of the sound event is then really important. 

\textbf{Scenario 2:}
The system must avoid confusion between classes but the reaction time is less crucial than in the first scenario.

Different PSDS sets of parameters need to be defined in order to reflect the particular needs of each scenario. For the DCASE 2021 Challenge Task 4, four parameters have been customized according to the different needs: Detection Tolerance criterion (DTC), Ground Truth intersection criterion (GTC), Cross-Trigger Tolerance criterion (cttc) and Cost of Cross-Triggers (CT) on the user experience ($\alpha_{CT}$). Table \ref{tab:PSDS} summarizes the PSDS values given the scenario. PSDS values are computed using 50 operating points (linearly distributed from 0.01 to 0.99). More information regarding PSDS can be found in Bilen et al.~\cite{PSDS:2020}.

The PSDS will be indicated throughout the paper as \textbf{PSDS\_1} when evaluated on scenario 1 and \textbf{PSDS\_2} when evaluated on scenario 2, regardless of the system. 

The systems analyzed on this paper have been selected according to the official ranking. The ranking criterion is the aggregation of PSDS\_1 and PSDS\_2. Each separate metric considered in the final ranking criterion is the best separate metric among all teams submission (each team is allowed 4 different submissions and PSDS\_1 and PSDS\_2 can be obtained by two different systems of the same team).

\begin{equation}
    \label{eq:PSDS}
    \mathrm{Ranking\ Score} = \overline{\mathrm{PSDS}\_1} + \overline{\mathrm{PSDS}\_2}
\end{equation}\
with $\overline{\mathrm{PSDS}\_1}$ and $\overline{\mathrm{PSDS}\_2}$ being the PSDS on scenario 1 and 2 normalized by the baseline PSDS on these scenarios, respectively. 
\begin{table}[t]
\centering
 \begin{tabular}{l|c|c}
  Parameter & Scenario 1 & Scenario 2 \\
  \toprule
  DTC & 0.7 & 0.1 \\
  GTC & 0.7 & 0.1 \\
  $\alpha_{CT}$ & 0 & 0.5 \\
  cttc & - & 0.3 \\
 \bottomrule
 \end{tabular}
 \caption{PSDS parameters of each scenario defined.}
 \label{tab:PSDS}
\end{table}

In this paper, we analyze different experiments to understand the systems behavior. The two different scenarios allows us to understand more in detail the possible adaptation of each system. The setup has been chosen in order to favor experiments on the systems behavior and their adaptation to different target scenario.

\subsection{Datasets}

\subsubsection{DESED dataset}
The dataset considered on this paper is the DESED dataset\footnote{\href{https://project.inria.fr/desed/}{https://project.inria.fr/desed/}} \cite{serizel:hal-02355573, turpault:hal-02160855}, which is the same as provided for the DCASE 2021 Challenge Task 4. 
It is composed of 10 seconds length audio clips either recorded in a domestic environment or synthesized to reproduce such an environment\footnote{For a detailed description of the DESED dataset and how it is generated the reader is referred to the original DESED article~\cite{turpault:hal-02160855} and DCASE 2021 task 4 webpage: \url{http://dcase.community/challenge2021}}.   
The synthetic part of the dataset is generated with Scaper \cite{salamon2017scaper}, a Python library for soundscape synthesis and augmentation. 
The foreground events (both target and non-target) are obtained from the Freesound Dataset (FSD50k) \cite{fonseca2020fsd50k}, while the background sounds are obtained from the SINS dataset (activity class “other”) \cite{dekkers2017sins} and TUT scenes 2016 development dataset \cite{mesaros2016tut}. The event co-occurences are computed on a set of strong annotations from Audioset~\cite{hershey2021benefit}.
More information regarding the generation of the DESED dataset can be found in Ronchini et al.~\cite{ronchini2021impact}. 


\subsection{Synthetic evaluation datasets}
The aim of this study is to investigate challenges related to real SED aspects such as target to non-target signal-to-noise ratio (TNTSNR), sound events localization in time and the impact of non-target sound events. Starting from the reference synthetic soundscapes evaluation set, eight different evaluation sets have been designed to specifically target these challenges. These additional evaluation datasets have been proposed to DCASE 2021 Task 4 participants together with the official evaluation set in order to be able to benchmark state-of-the-art systems on these particular aspects.

\subsubsection{Reference synthetic soundscapes evaluation set}
\label{sub:refset}

The synthetic 2021 evaluation set is composed of 1000 audio clips.
In the context of the challenge, this subset is used for analysis purposes. We will refer to it as \textbf{synth}, being the reference evaluation synthetic set. For consistency, on the figures on the paper the set \textbf{synth} will be referred to as  \textbf{TNTSNR\_0}, being 0 the TNTSNR. It contains target and non-target events distributed between the different audio clips according to pre-calculated co-occurrences \cite{ronchini2021impact}. Two additional versions of the \textbf{synth} set have been generated, \textbf{TNTSNR\_inf} (only target sound events) and \textbf{synth\_ntg} (only non-target sound events). 

\subsubsection{Synthetic set with varying TNTSNR}
\label{sub:TNTSNRset}

With the aim of studying what would be the impact of varying the TNTSNR on the system performance, three different versions of \textbf{synth} have been generated. The SNR of non-target events have been decreased by 5 dB, 10 dB and 15 dB compared to their original value. The original SNR of the sound events is randomly selected between 6 dB and 30 dB, so the more we decrease the SNR, the less the sound will be audible, with some of the events that will not be audible at all. These subsets will be subsequently referred to as \textbf{TNTSNR\_5}, \textbf{TNTSNR\_10}, \textbf{TNTSNR\_15}.

\subsubsection{Synthetic set with varying onset time}
\label{sub:onsetset}

A subset of 1000 soundscapes is generated with a uniform sound event onset distribution and only one event per soundscape.
Three variants of this subset have been generated
with the event onset shifted in time.
The sound event onsets are located between 250~ms and 750~ms in the first version, between 5.25~s and 5.75~s in the second version and between 9.25~s and 9.75~s in the third version.
These subsets will be hereafter referred to as \textbf{500ms}, \textbf{5500ms} and \textbf{9500ms}, respectively. These subsets are designed to study the sensibility of the SED segmentation to the sound event localization in time. A similar experiment has already been conducted for DCASE 2019 and DCASE 2020 Task 4~\cite{ Serizel2020_ICASSP, turpault2021sound}. The main purpose here is to analyze whether the systems have improved on this particular aspect.

\begin{figure}[t]
\includegraphics[width=0.5\textwidth]{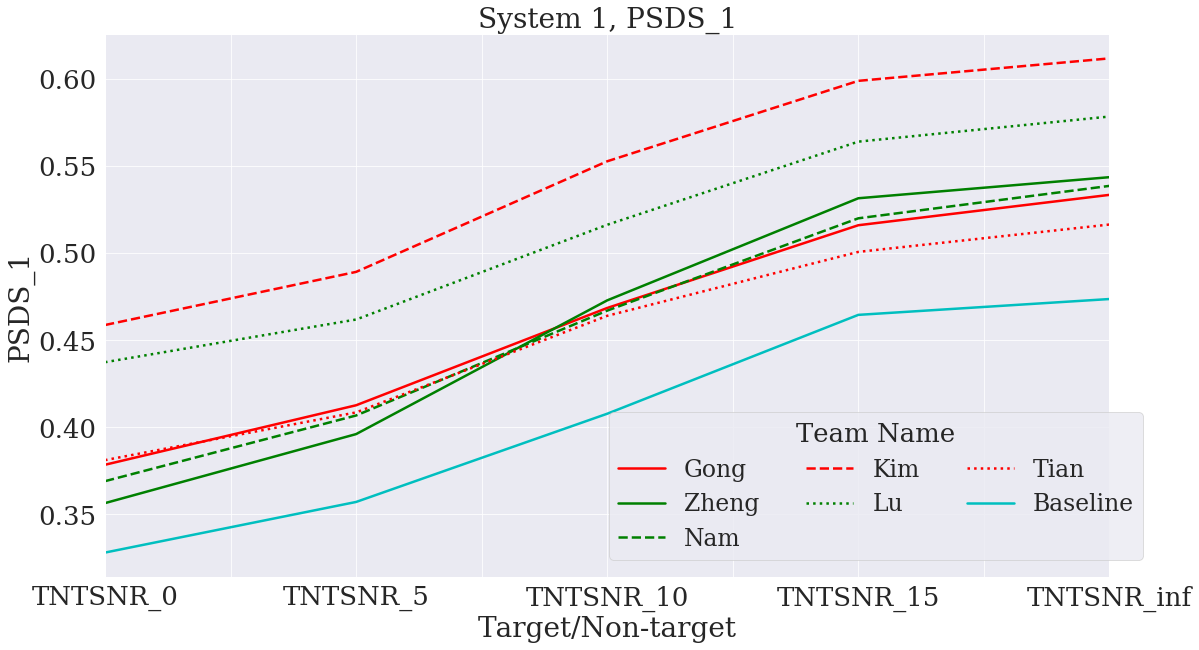}
\caption{PSDS\_1 results for systems selected for scenario 1, evaluated with reference evaluation set and sets with varying TNTSNR.}
\label{fig:sis1p1}
\end{figure}

\section{Impact of the evaluation scenario}
\label{sec:evalscen}
One novelty of the DCASE Challenge 2021 Task 4 is the introduction of two evaluation scenarios. As described in Section \ref{subsec:setup}, one scenario is strict in terms of temporal segmentation (scenario 1) while the other is more permissive in terms of temporal segmentation but systems must avoid confusion between classes (scenario 2). Since the participants were allowed to present up to 4 different systems, we first analyze whether or not the participants submitted specific systems for the scenarios.

Table \ref{tab:rank} presents the PSDS scores obtained on the official evaluation set for both scenarios. In particular, the table reports the results of the 9 highest-ranked teams and the baseline. For each team, the PSDS for the two systems selected for the ranking score are reported.
For three participants, the systems that perform best on scenario 1 and scenario 2 have different behaviors ~\cite{Zheng2021,Lu2021,Nam2021}. Zheng et al. adjust the sigmoid temperature parameter to obtain soft or sharp detection output depending on the scenario~\cite{Zheng2021}. Lu et al. use a convolutional nerual network (CNN) model on scenario 1 and a conformer on scenario 2~\cite{Lu2021}.  Nam et al. propose to train a weak SED system for scenario 2, primarily focusing on event classification~\cite{Nam2021}. For some participants, the best system on scenario 1 is also the best system on scenario 2~\cite{Kim2021,Tian2021,Gong2021}. In the remainder of the paper we focus on the six submissions listed above, comparing their performance to those obtained with the baseline. Table~\ref{tab:rank} also reports performance obtained by other participants for which the difference between the best system on scenario 1 and the best system on scenario 2 is marginal (the difference is mainly due to a different system adjustment)~\cite{Ebbers2021,Wang2021,Yu2021}. 

\begin{figure}[t]
\includegraphics[width=0.5\textwidth]{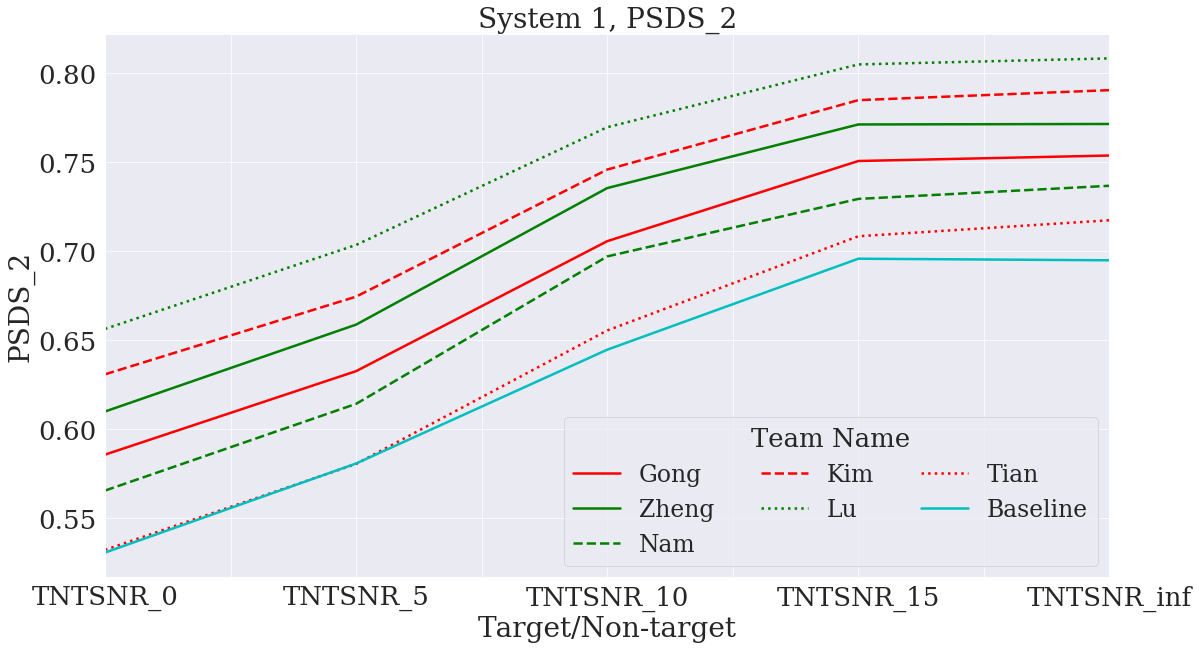}
\caption{PSDS\_2 results for systems selected for scenario 1, evaluated with reference evaluation set and sets with varying TNTSNR.}
\label{fig:sis1p2}
\end{figure}

\section{ANALYSIS ON THE SYNTHETIC DATASETS}
\label{sec:analysis}

This section compares the different submissions evaluated on the different synthetic datasets described in Section~\ref{sec:dataset} in order to investigate the impact of the TNTSNR (Section \ref{subsec:tntsnr}), the sound event localization within the clip (Section \ref{subsec:onset}) and the impact of non-target sound events when considered alone (Section  \ref{subsec:nontg}).

The objective here is to highlight the different impact of using systems that are fine-tuned according to the different scenarios compared to using a general system. Therefore, all the plots in this paper use different colors to identify the teams. The performance for the teams using different systems for the two scenarios~\cite{Zheng2021,Lu2021,Nam2021} are represented with green lines, while the performance for the teams using the same system for both scenarios~\cite{Kim2021,Tian2021,Gong2021} are presented in red. The cyan line represents the baseline performance. 

\subsection{Impact of TNTSNR}
\label{subsec:tntsnr}
\begin{figure}[t]
\includegraphics[width=0.5\textwidth]{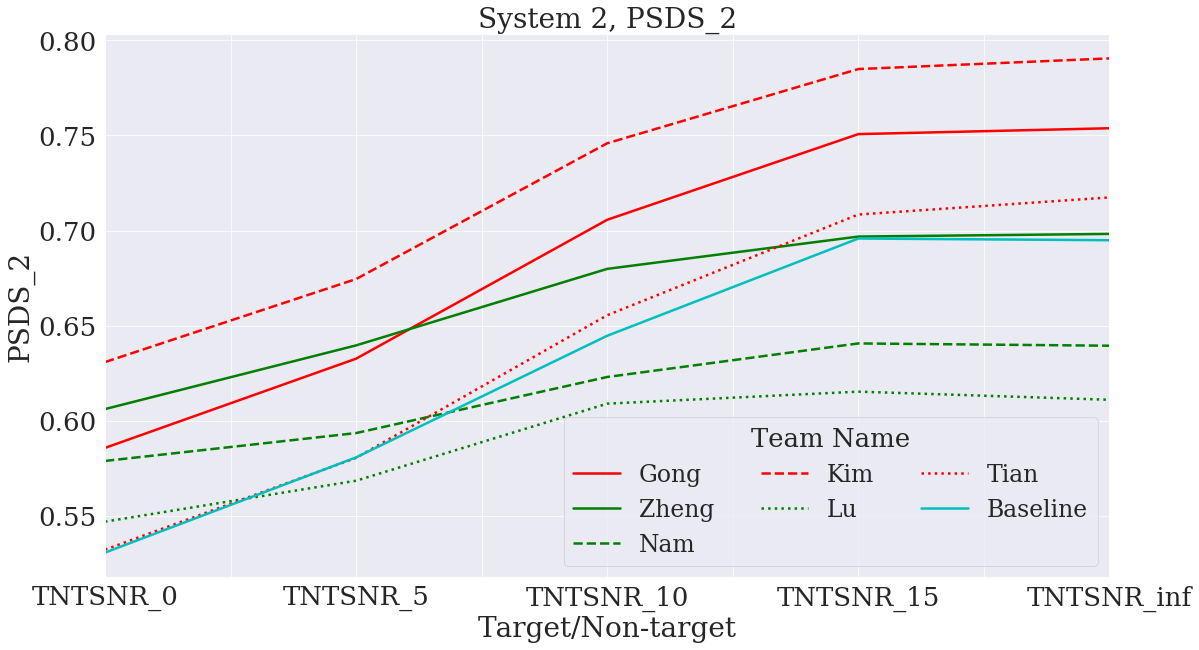}
\caption{PSDS\_2 results for systems selected for scenario 2, evaluated with reference evaluation set and sets with varying TNTSNR.}
\label{fig:sis2p2}
\end{figure}

Figures \ref{fig:sis1p1}, \ref{fig:sis1p2} and \ref{fig:sis2p2} report the results obtained with the submissions evaluated on the dataset with varying TNTSNR (see also Sections \ref{sub:refset} and \ref{sub:TNTSNRset}). 
Figures \ref{fig:sis1p1} and \ref{fig:sis1p2} show the performance of the systems selected on scenario 1, reporting both PSDS\_1 and PSDS\_2 whereas Figure \ref{fig:sis2p2} reports the PSDS\_2 for the submissions selected on scenario 2. 

As can be observed from Figures \ref{fig:sis1p1} and \ref{fig:sis1p2}, all the submissions (selected on scenario 1 emphasizing the temporal segmentation) perform better when only target events are present in the evaluation set, with the performance that consistently decreases with the TNTSNR getting lower. This confirms the results observed in previous work~\cite{turpault:hal-02891700}. The drop between the best performance (without non-target sound events) and the worst performance (TNTSNR\_0) is similar with both metrics (Figure~\ref{fig:sis1p1} and~\ref{fig:sis1p2}) regardless of the systems. This could indicate that the TNTSNR has little effect on the segmentation performance of the systems.

Figure \ref{fig:sis2p2} shows the PSDS\_2 of the submissions selected for scenario 2 (relaxed time segmentation constraint). It is interesting to notice that the results obtained with systems that are tailored to provide coarse segmentation  are more robust to different TNTSNR. In fact, with respect to this scenario, which emphasizes sound event class prediction over time segmentation, they evince a relatively smaller drop between the best performance (without non-target sound events) and the worst performance (TNTSNR\_0) compared to the general systems. 
This tends to indicate that strong sound event classification capabilities are particularly important when considering robustness to varying TNTSNR. 

\subsection{Impact of time localization of the original event}
\label{subsec:onset}

\begin{figure}[t]
\includegraphics[width=0.5\textwidth]{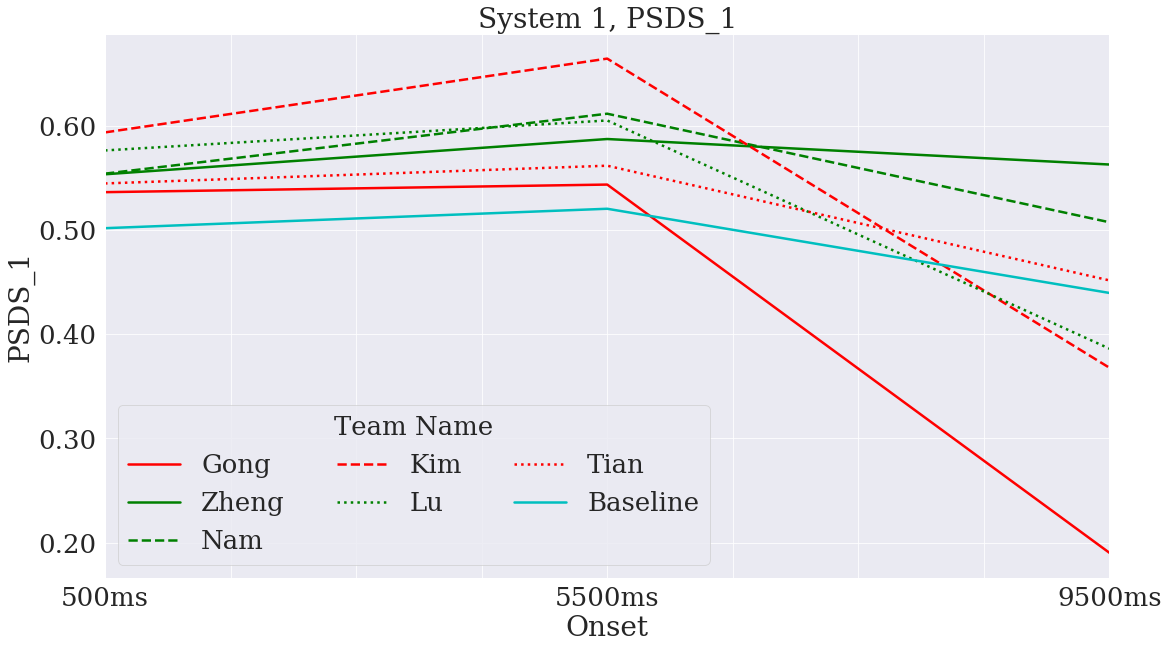}
\caption{PSDS\_1 results for systems selected for scenario 1, evaluated with synthetic set with varying onset time.}
\label{fig:onset}
\end{figure}

Figure \ref{fig:onset} shows the PSDS\_1 for the submissions selected on scenario 1. In this experiment, we focus on systems selected for scenario 1 as the problem considered in this part of the analysis is related to time localization. We evaluate the systems using the evaluation sets described in Section \ref{sub:onsetset}. This experiment has already been proposed for DCASE 2019 and DCASE 2020 submissions \cite{Serizel2020_ICASSP, turpault2021sound}. Previous analyses showed that performance consistently drop when the onsets of the sound events are located towards the end of the clips~\cite{turpault2021sound}. 
Similar performance trends are obtained with general
systems~\cite{Kim2021,Tian2021,Gong2021}, while systems that have been adapted for scenario 1~\cite{Zheng2021,Lu2021,Nam2021} generally show attenuated performance drop towards the end of the clips.
In particular, the approaches proposed by Zheng et al.~\cite{Zheng2021} and Nam et al.~\cite{Nam2021} have robust performance regardless of the onset timing. According to the systems description, this improvement seems to be related to the use of specific data augmentation methods.

\subsection{Impact of non-target sound events}
\label{subsec:nontg}
In this last experiment, we investigate the impact of non-target sound events on the systems outputs. Similarly to the study reported on Ronchini et al.~\cite{ronchini2021impact}, 
we are interested in studying whether false positive events can be triggered by non-target events and identify which ones. In order to do so, the systems have been evaluated with the \textbf{synth\_ntg} evaluation set, described in Section \ref{sub:refset}.

Table~\ref{tab:ntg_events} presents the number of target events detected by the systems on clips that do not contain any target event. The results are split depending on the average length of the target classes detected:
\textit{Alarm bell ringing, Cat, Dishes, Dog and Speech} are considered as short event classes while \textit{Blender, Running water, Electric shaver toothbrush, Frying and Vacuum cleaner} are considered as long event classes. Systems tend to spuriously predict short events more than long events. This is particularly true for systems that are tailored to have a fine segmentation (Zheng\_SED\_1, Lu\_SED\_1, Nam\_SED\_2). This sensitivity probably has to be taken into account when designing systems with fine segmentation.

\begin{table}[ht]
\centering
 \begin{tabular}{l|c|c|c}
  \toprule
  Submission code & All events & Short events & Long events \\
  \midrule 
  Zheng\_SED\_1 & 721 & 665 & 56 \\
  Zheng\_SED\_3 & 448 & 392 & 56\\
  
  Lu\_SED\_1 & 781 & 719 & 62 \\
  Lu\_SED\_3 & 282 & 225 & 57 \\
  
  Nam\_SED\_2 & 1098 & 1044 & 54 \\ 
  Nam\_SED\_4 & 500 & 434 & 66 \\
  
  \midrule
  Baseline & 831 & 697 & 134 \\ 
  
 \bottomrule
 \end{tabular}
 \caption{Number of non-target events detected by the systems when evaluated with \textbf{sytnh\_ntg}.}
 \label{tab:ntg_events}
\end{table}

\section{CONCLUSIONS}
\label{sec:typestyle}
This paper presents a benchmark of state-of-the-art Sound Event Detection systems (submissions to DCASE 2021 Challenge Task 4) on evaluation sets composed of synthetic soundscapes designed to target specific challenges of the Sound Event Detection task. The main challenges addressed in this paper are the impact of varying target to non-target signal-to-noise ratio, the impact of time localization of the sound event and the impact of non-target sound events. We observe that systems that are tailored for a fine time segmentation are generally more robust to the event localization within the clips but can also be more sensitive to false alarm triggered by non-target events. On the other end, systems that are tailored for coarse time segmentation generally provide an event classification that is more robust to low TNTSNR.




\vfill\pagebreak
\clearpage

\bibliographystyle{IEEEbib}
\bibliography{refs,technical_reports_task4}

\end{document}